\documentclass[usenatbib]{article}
\usepackage{graphicx}
\usepackage{amsmath}
\usepackage{natbib}
\usepackage{amssymb}

\usepackage{sidecap}

\usepackage[breaklinks,colorlinks,urlcolor=blue,citecolor=blue,linkcolor=blue]{hyperref}
\usepackage{hyperref}
\newcommand       \procspie          {Procspie}
\newcommand       \apj          {ApJ}
\newcommand       \apjl         {ApJL}
\newcommand       \aap          {A\&A}

\newcommand       \prc          {PRC}
\newcommand       \nat          {Nature}
\newcommand       \mnras        {MNRAS}
\newcommand       \pasp      {PASP}

\newcommand       \prd      {Phys.~Rev.~D.~}

\newcommand       \araa      {ARA\&A}

\def\LIGO{GW170817}

\def\LIGO{GW170817}

\begin{document}

\newcommand{\be}{\begin{equation}}
\newcommand{\ee}{\end{equation}}

\title{Welcome to the Multi-Messenger Era! \\ Lessons from a Neutron Star Merger and the Landscape Ahead}

\author{Brian D.~Metzger\\ 
Columbia Astrophysics Laboratory\\
Columbia University\\
bmetzger@phys.columbia.edu}

\maketitle

\abstract{The discovery by Advanced LIGO/Virgo of gravitational waves from the binary neutron star merger \LIGO, and subsequently by astronomers of transient counterparts across the electromagnetic spectrum, has initiated the era of ``multi-messenger astronomy".   Given the slew of papers appearing on this event, I thought it useful to summarize the electromagnetic discoveries in the context of theoretical counterpart models and to present personal views on the major take-away lessons and outstanding new questions from this watershed event.  The weak burst of gamma-rays discovered in close time coincidence with \LIGO, and potential evidence for a more powerful off-axis relativistic jet (initially beamed away from our line of sight) via the delayed rise of a non-thermal X-ray and radio orphan afterglow, provides the most compelling evidence yet that cosmological short gamma-ray bursts originate from binary NS mergers.  The luminosity and colors of the early optical emission discovered within a day of the merger agrees strikingly well with original predictions of \citet{Metzger+10} for ``kilonova" emission powered by the radioactive decay of $r$-process nuclei, the binary NS merger origin of which was initially proposed by \citet{Lattimer&Schramm74}.  The transition of the spectral energy distribution to near-infrared wavelengths on timescales of days matches the predictions by \citet{Barnes&Kasen13} and \citet{Tanaka&Hotokezaka13} if a portion of the ejecta contains heavy $r$-process nuclei with higher opacities due to the presence of lanthanides.  The ``blue" and ``red" ejecta components may possess distinct origins (e.g.~dynamical ejecta versus accretion disk outflows), with key implications for the physics of the merger and the properties of neutron stars.  I outline the diversity in the counterpart emission expected from additional mergers$-$observed with a range of different binary masses and viewing angles$-$discovered in the years ahead as LIGO/Virgo approach design sensitivity and NS mergers are discovered as frequently as once per week.}

\section{A Big Reveal from the Cosmos}

When a neutron star (NS) binary coalesces into a single object following a prolonged inspiral driven by gravitational wave (GW) radiation, the outcome is a prodigious collision which releases mass and energy into the surrounding environment on a timescale as short as milliseconds.  The merger aftermath was predicted to be accompanied by a diverse range of thermal and non-thermal electromagnetic (EM) counterparts from radio to gamma-ray frequencies (e.g.~\citealt{Bloom+09,Metzger&Berger12,Piran+13,Rosswog15,Fernandez&Metzger16}).  The discovery of the first GW chirp from a binary NS merger \LIGO~by the Advanced LIGO and Virgo collaboration \citep{LIGO+17DISCOVERY}, and its subsequent localization to a host galaxy at a distance of only $\approx$ 40 Mpc (e.g.~\cite{LIGO+17CAPSTONE} and references therein), provided astronomers with a golden opportunity to test theoretical predictions for the EM and nucleosynthetic signatures of these events, as established by the work of astrophysicists over the last 40 years. 

The discovery and announcement of \LIGO~was followed by the most ambitious (and emotionally charged) campaign of EM follow-up observations ever conducted \citep{LIGO+17CAPSTONE}.  Observations covered the gamut of frequencies, including radio/microwave (e.g.~\citealt{Mooley+17,Murphy+17,Hallinan+17,Alexander+17}), infrared (e.g.~\citealt{Chornock+17,Levan&Tanvir17,Kasliwal+17}), optical/UV (e.g.~\citealt{Coulter+17,Allam+17,Yang+17,Arcavi+17,Kilpatrick+17,McCully+17,Pian+17,Arcavi+17b,Tanvir&Levan17,Evans+17,Lipunov+17,Cowperthwaite+17,Smartt+17,Shappee+17}), X-ray \citep{Troja+17,Margutti+17,Haggard+17,Fong+17},  gamma-ray (e.g.~\citealt{Goldstein+17,Savchenko+17,LIGO+17FERMI,Verrechia+17}), and even neutrinos (\citealt{LIGO+17NEUTRINOS}).  The full range of observational references are summarized in \citet{LIGO+17CAPSTONE}.

\begin{SCfigure}
\centering
\includegraphics[width=0.6\textwidth]
{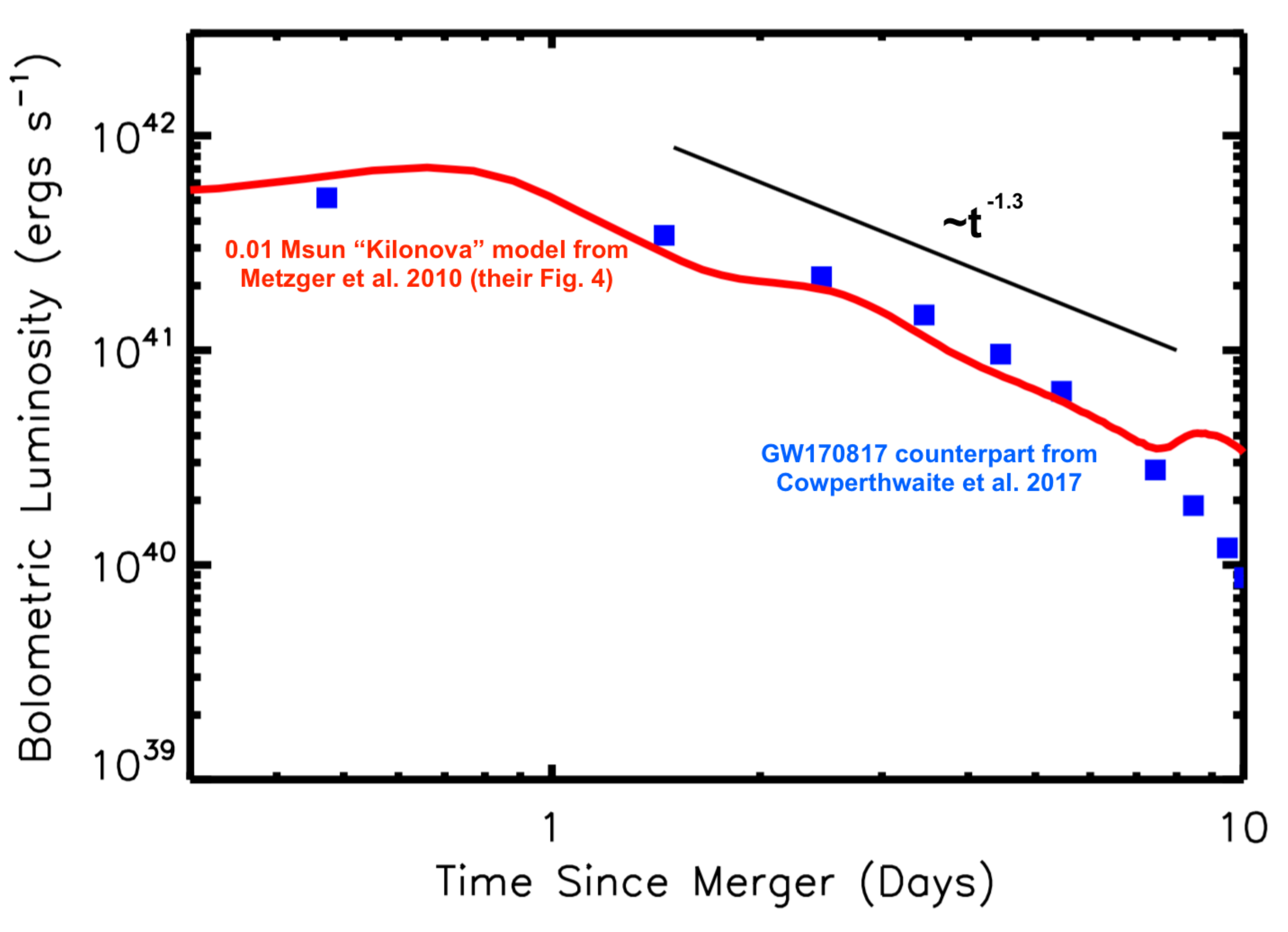}
% un-comment the following line to include your fig1b.ps postscript file:
\caption{\footnotesize  Bolometric light curve of the optical/infrared counterpart of \LIGO~(blue squares) from multi-band photometry \citep{Cowperthwaite+17} compared to the fiducial model of \citet{Metzger+10} (red line; their Fig.~4) for ``kilonova" emission powered by the radioactive decay of $10^{-2}M_{\odot}$ of $r$-process matter expanding at $v = 0.1$ c, assuming complete thermalization of the radioactive decay products.  Shown above for comparison is a line with the approximate power-law decay  $\propto t^{-1.3}$ for $r$-process heating \citep{Metzger+10,Hotokezaka+17}.  The true ejecta mass required to explain the data exceeds 0.01$M_{\odot}$ by a factor of several (Table \ref{table:BNS}) because the actual thermalization efficiency is less than unity \citep{Barnes+16,Rosswog+17}.  The observed color evolution of the transient from optical to near-infrared wavelengths can also only be understood by accounting for the details of the ejecta structure and the different opacities of light and heavy $r$-process nuclei ($\S\ref{sec:KN}$ for details).   }
\label{fig:Lbol}
\vspace{-0.2cm}
\end{SCfigure}

Often in astronomy, hints of the underlying truth about a phenomenon build up only gradually as the capabilities of telescopes incrementally improve; and even once a consensus opinion is reached, it is often the product of several pieces of indirect evidence.  \LIGO~represents a sharp departure from this rule, as  LIGO/Virgo transported us, in one quantum leap, directly from the dark into the light (the ``Big Reveal"), albeit a leap that theorists had long anticipated and given unusually extensive consideration to, despite the lack of observational guidance.  As information on the discovery percolated in, I was overtaken by the degree to which the optical and infrared transient being observed agreed with those predicted by myself and colleagues, such as work I led in 2010 (Fig.~\ref{fig:Lbol}).  Seeing Nature agree so well with our basic ideas is a triumph for astrophysics theory.

\begin{SCfigure}
\centering
\includegraphics[width=0.6\textwidth]
{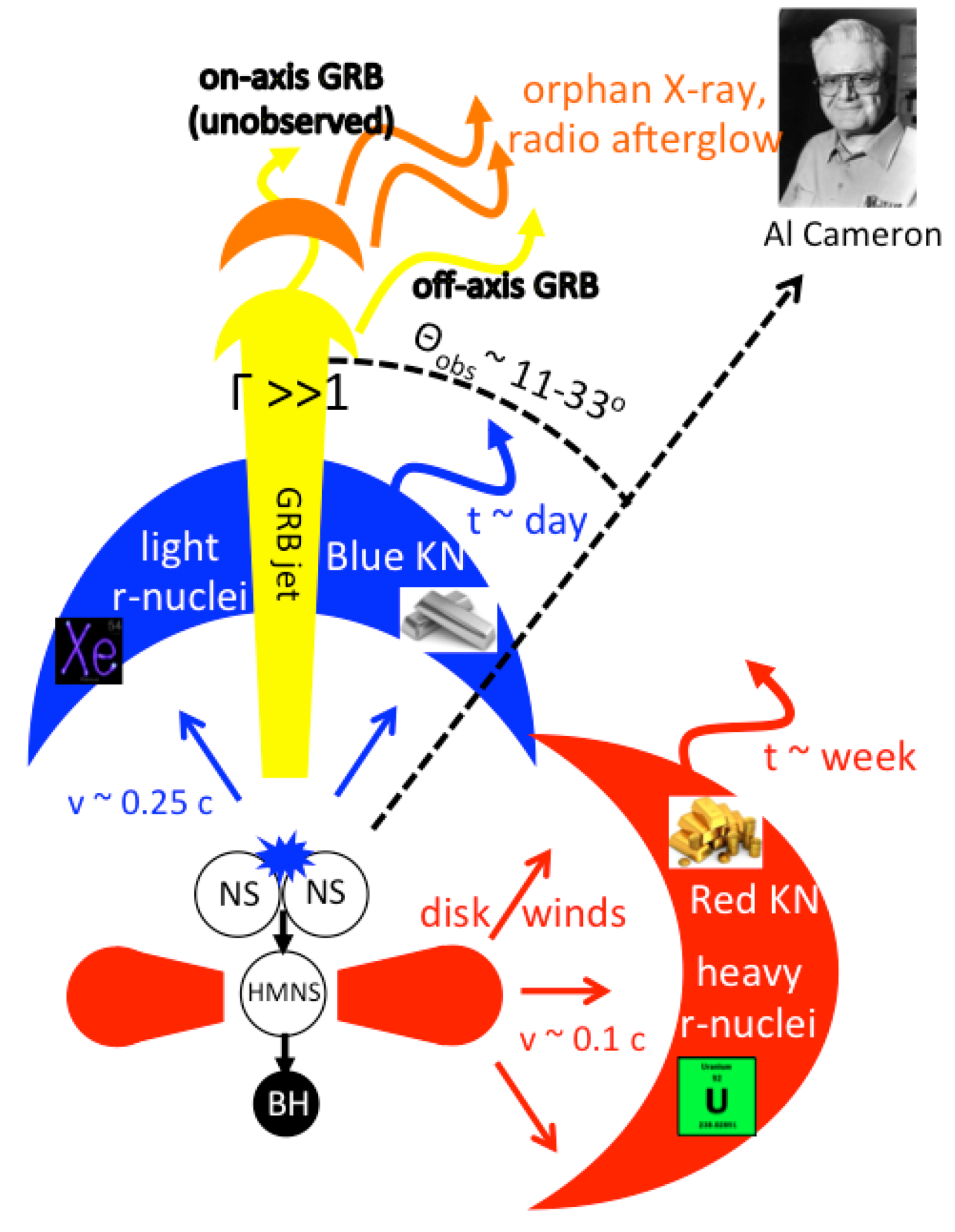}
% un-comment the following line to include your fig1b.ps postscript file:
\caption{\footnotesize Scenario for the EM counterparts of \LIGO, as viewed by the observer (Al Cameron) from the inferred binary inclination angle $\theta_{\rm obs} \approx 0.2-0.5$ \citep{LIGO+17H0}, as motivated by interpretations presented in several papers (e.g.~\citealt{Cowperthwaite+17,Kasen+17,Nicholl+17,Chornock+17,Fong+17,Kasen+17,Margutti+17,LIGO+17FERMI}).
{\bf Timeline:} (1) Two NSs with small radii $\lesssim 11$ km and comparable masses ($q \approx 1$) coalesce.  The dynamical stage of the merger ejects only a small mass $\lesssim 10^{-2}M_{\odot}$ in equatorial tidal ejecta, but a larger quantity $\approx 10^{-2}M_{\odot}$ of $Y_{e}>0.25$ matter into the polar region at $v \approx 0.2-0.3$ c, which synthesizes exclusively light $r$-process nuclei (e.g.~xenon and silver); (2) The merger product is a meta-stable hypermassive NS, which generates a large accretion torus $\sim 0.1M_{\odot}$ as it sheds its angular momentum and collapses into a BH on a timescale of $\lesssim 100$ ms; (3) The torus-BH powers a collimated GRB jet, which burrows through the polar dynamical ejecta on a timescale of $\lesssim$ 2 s; (4) Gamma-rays from the core of the GRB jet are relativistically beamed away from our sight line, but a weaker GRB is nevertheless observed from the off-axis jet or the hot cocoon created as the jet breaks through the polar ejecta; (5) On a similar timescale, the accretion disk produces a powerful wind ejecting $\approx 0.04M_{\odot}$ of $Y_{e} \lesssim 0.25$ matter which expands quasi-spherically at $v \approx 0.1$ c and synthesizes also heavy $r$-process nuclei such as gold and uranium; (6) After several hours of expansion, the polar ejecta becomes diffusive, powering $\sim$ visual wavelength (``blue") kilonova emission lasting for a few days; (7) over a longer timescale $\approx$ 1 week, the deeper disk wind ejecta becomes diffusive, powering red kilonova emission; (8) the initially on-axis GRB jet decelerates by shocking the ISM, such that after $\approx 2$ weeks its X-ray and radio synchrotron afterglow emission rises after entering the observer's causal cone. }
\label{fig:schematic}
\vspace{-0.2cm}
\end{SCfigure}

Given the slew of observational and interpretation papers appearing on this topic over just a few days, I thought it useful to review briefly, in one place, theoretical models for the EM counterparts of binary NS mergers in the context of the \LIGO~discovery.  I start by describing the thermal kilonova emission coming from the mildly-relativistic merger ejecta ($\S\ref{sec:kilo}$) and then discuss non-thermal emission from the ultra-relativistic gamma-ray burst (GRB) jet ($\S\ref{sec:GRB}$).  Figure \ref{fig:schematic} summarizes a reasonable guess for the origin of the different EM counterparts observed following~\LIGO.  In $\S\ref{sec:takeaways}$, I draw major take-away lessons from the first binary NS merger, and use them to motivate new questions for scrutiny as the sample of EM/GW events grows over the next several years.  Many of the interpretations presented result from interaction with the observational groups in which I collaborated, particularly the Dark Energy Camera (DECam) group, and I encourage the reader to consult these works for in-depth analysis of these data.

\begin{table}
\caption{Key Properties of \LIGO \label{table:BNS}}

\begin{tabular}{ccc}

Property & Value & Reference  \\

\hline

Chirp mass, $\mathcal{M}$ (rest frame) & 1.188$^{+0.004}_{-0.002}M_{\odot}$  & 1 \\
First NS mass, $M_{\rm 1}$ & $1.36-1.60M_{\odot}$ (90\%, low spin prior) & 1\\
Second NS mass, $M_{\rm 2}$ &  $1.17-1.36M_{\odot}$  (90\%, low spin prior) & 1\\
Total binary mass, $M_{\rm tot} = M_{1}+M_{2}$ & $\approx 2.74^{0.04}_{-0.01}M_{\odot}$ &  1 \\
Observer angle relative to binary axis, $\theta_{\rm obs}$ & $11-33^{\circ}$ (68.3\%) &  2  \\
Blue KN ejecta ($A_{\rm max} \lesssim 140$) & $\approx 0.01-0.02M_{\odot}$ & e.g., 3,4,5 \\
Red KN ejecta ($A_{\rm max} \gtrsim 140$) & $\approx 0.04M_{\odot}$ & e.g., 3,5,6 \\ 
Light $r$-process yield ($A \lesssim 140$) & $\approx 0.05-0.06M_{\odot}$ & \\
Heavy $r$-process yield ($A \gtrsim 140$) & $\approx 0.01M_{\odot}$ & \\ 
Gold yield & $\sim 100-200 M_{\oplus}$ & 8 \\
Uranium yield & $\sim 30-60M_{\oplus}$ & 8 \\
Kinetic energy of off-axis GRB jet & $10^{49}-10^{50}$ erg & e.g., 9, 10, 11, 12\\
ISM density & $10^{-4}-10^{-2}$ cm$^{-3}$ & e.g., 9, 10, 11, 12 \\
\hline
\end{tabular} 
\newpage
(1) \citealt{LIGO+17PARAMS}; (2) depends on Hubble Constant, \citealt{LIGO+17H0}; (3) \citealt{Cowperthwaite+17}; (4) \citealt{Nicholl+17}; (5) \citealt{Kasen+17}; (6) \citealt{Chornock+17}; (8) assuming heavy $r$-process ($A > 140$) yields distributed as solar abundances \citep{Arnould+07}; (9)\citealt{Margutti+17}; (10) \citealt{Troja+17}; (11) \citealt{Fong+17}; (12) \citealt{Hallinan+17}
\end{table}

\section{Kilonovae and the Origin of the Heaviest Elements}
\label{sec:kilo}

The optical/infrared transient following \LIGO~is fully consistent with being powered by the radioactive decay of nuclei synthesized in the NS merger ejecta.  Here, I review the history of models for the $r$-process in binary NS mergers and the expected sources of mass ejection in these events based on numerical simulations (\S\ref{sec:rprocess}).  I then describe the historical development of kilonova models ($\S\ref{sec:KN}$) in the context of their expected timescales, luminosities and colors; particular emphasis is placed on the distinction between the emission from ejecta containing light versus heavy $r$-process nuclei.  Within this framework, in \S\ref{sec:KNinterp} I summarize our interpretation for the kilonova from~\LIGO, and the resulting implications for the fate of the merger remnant and the properties of NSs more broadly.

\subsection{Mass Ejection in Binary NS Mergers and the $r$-Process}
\label{sec:rprocess}

\begin{figure}[!t]
\includegraphics[width=1.0\textwidth]{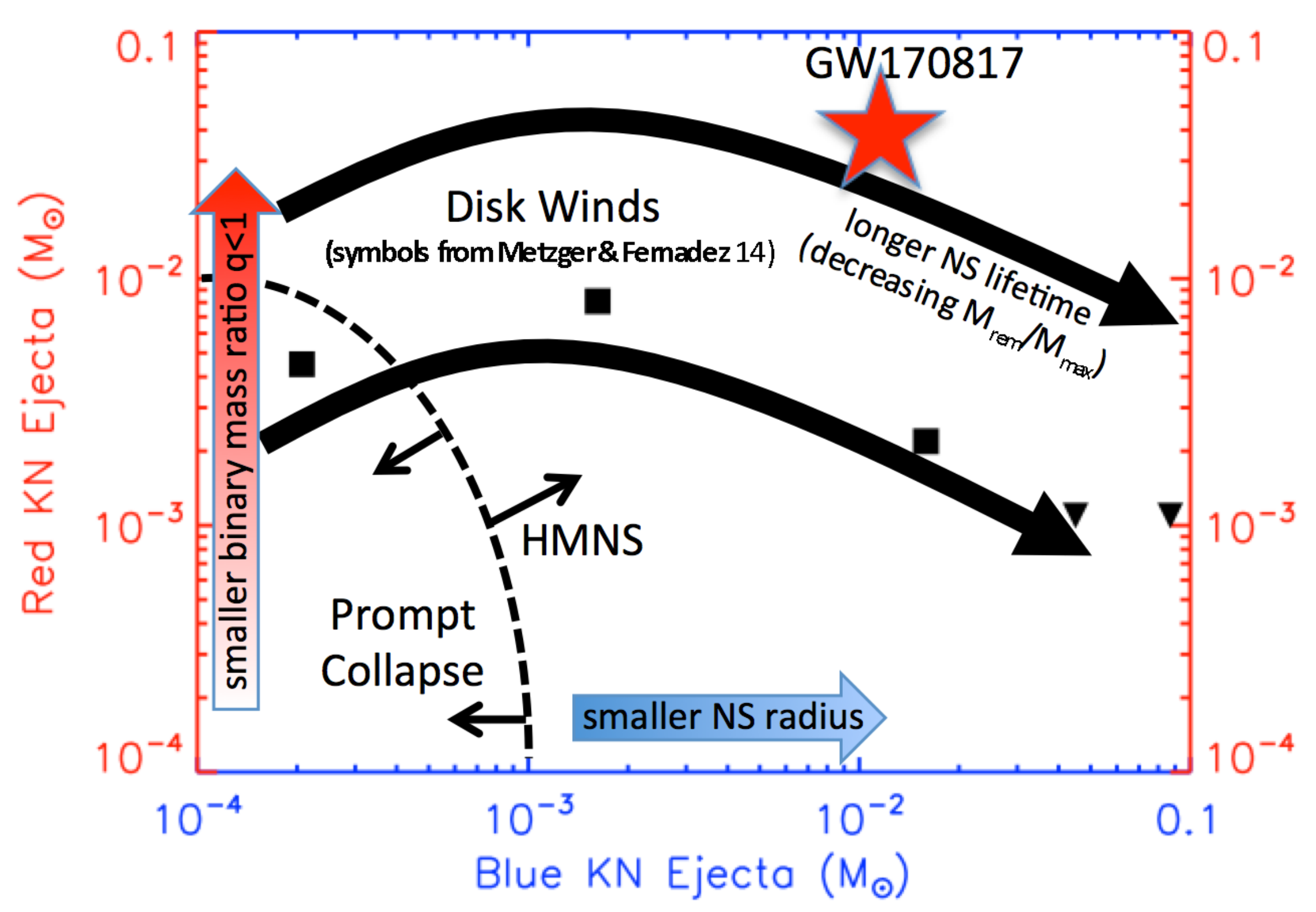}
\hspace{0.0cm}
\caption{\footnotesize Quantity of lanthanide-free (light $r$-process; ``blue KN") ejecta and lanthanide-bearing (heavy $r$-process; ``red KN") ejecta from a binary NS merger and its dependence on the properties of the binary (remnant mass $M_{\rm rem}$, NS radius, and maximum mass NS $M_{\rm max}$) in comparison to those inferred from the blue and red kilonova emission of~\LIGO~(e.g.~\citealt{Cowperthwaite+17,Nicholl+17,Chornock+17}).  The amount of low-$Y_{e}$ (red) tidal tail ejecta increases for more asymmetric binaries (decreasing binary mass ratio $q = M_{2}/M_{1} < 1$), while the amount of high-$Y_{e}$ (blue) shock-heated ejecta ejected dynamically into the polar regions is larger if the colliding NSs possess smaller radii.  For a massive binary with a high ratio of $M_{\rm rem}/M_{\rm max}$, the merger product promptly collapses to a BH, producing little blue shock-heated ejecta.  The dependence of the disk wind ejecta composition, as approximately delineated as the region between the black arrows, is more complex and depends on the lifetime of the hyper- or supra-massive NS merger remnant, which increases with decreasing $M_{\rm rem}/M_{\rm max}$ (\citealt{Metzger&Fernandez14,Perego+14,Kasen+15}).}
\label{fig:ejecta}
\end{figure}

Roughly 60 years ago, \citet{Burbidge+57} and \citet{Cameron57} recognized that approximately half of the elements in the Galaxy heavier than iron must have been produced in an environment in which the density of free neutrons was so high that neutron captures on nuclei proceed much faster than $\beta-$decays.  Since that time, however, while the astrophysical sites of most of the other nucleosynthesis channels identified in these pioneering works have been identified, the origin of the rapid neutron capture process (``$r$-process", for short) has remained an enduring mystery.  Hot outflows from the newly-formed proto-neutron stars created in core collapse supernovae were at one time considered the most promising contender.  However, this model is known to exhibit large theoretical difficulties (e.g.~\citealt{Qian&Woosley96}), and several lines of evidence in recent years have pointed towards an $r$-process source which is much rarer than standard supernovae \citep{Wallner+15,Hotokezaka+15,Ji+16,Macias&RamirezRuiz16}

\citet{Lattimer&Schramm74} proposed that the coalescence of a binary system consisting of a NS and a stellar mass black hole (BH) would provide a promising source of neutron-rich ejecta conducive to the $r$-process with a very low electron fraction $Y_{e} = n_{\rm p}/(n_{\rm n} + n_{\rm p})$, where $n_{\rm p}$ and $n_{\rm n}$ are the densities of protons and neutrons, respectively.  Following the ejection of NS matter through the outer binary Lagrange points by tidal forces, its rapid decompression from nuclear densities would naturally result in the formation of heavy nuclei through neutron capture \citep{Lattimer+77,Meyer89}.  \citet{Symbalisty&Schramm82} and \citet{Eichler+89} proposed that a similar mechanism of mass ejection could occur from merging compact binaries consisting of two NSs.  The first numerical simulations of binary NS mergers showing tidal mass ejection followed (\citealt{Davies+94,Ruffert+97,Rosswog+99}), with subsequent work establishing that the $r$-process of this extremely neutron-rich matter ($Y_{e} \lesssim 0.1-0.2$) would result in an heavy element abundance pattern broadly consistent with that in the solar system \citep{Freiburghaus+99,Goriely+11}.  Tidally-ejected matter expands away from the merger site primarily in the equatorial plane of the binary at velocities $\sim 0.2-0.3$ c, and its quantity $\sim 10^{-4}-10^{-2}M_{\odot}$ is a sensitive decreasing function of the binary mass ratio $q = M_{2}/M_{1} < 1$ (e.g.~\citealt{Rosswog+99,Hotokezaka+13}), i.e.~more asymmetric mergers eject greater mass tidally. 

In addition to the tidal ejecta, contemporary numerical simulations have established a separate ejecta source originating from the interface between the merging stars and emerging into the high latitude polar region \citep{Oechslin&Janka06,Hotokezaka+13,Bauswein+13}.  Heating due to shocks and neutrino-irradiation promote weak interactions (e.g.~$n \nu_{e} \rightarrow p e^{-}$, $e^{+}n \rightarrow p \bar{\nu}_{e}$) which raise the electron fraction of the polar ejecta to values $Y_{e} \gtrsim 0.25$ well above its initial composition in the neutron star interior \citep{Wanajo+14,Goriely+15,Sekiguchi+16,Radice+16}.  Though relatively independent of the binary mass ratio, the quantity of shock-heated polar ejecta instead depends sensitively on the NS radius (see below) and the lifetime of the compact remnant created during the merger.  If the baryonic mass of the binary (and thus of its compact central remnant, $M_{\rm rem}$) exceeds the maximum mass of a neutron star, $M_{\rm max}$, by a factor of $f = 1.3-1.6$ (the precise value depending on the EOS; \citealt{Bauswein+13b}b), then the merger product  promptly collapses to a BH with little or no polar dynamical ejecta \citep{Shibata&Taniguchi06}.  On the other hand, if $M_{\rm rem} \lesssim f M_{\rm max}$ then the merger product is a hyper- or supra-massive NS, which is at least temporarily stable to collapse due to its rapid rotation.  The quantity of polar dynamical ejecta in this case exceeds the prompt collapse case, varying from $\sim 10^{-3}-10^{-2}M_{\odot}$, depending most sensitively on the radii of the NSs; a more compact NS results in the collision occurring deeper in the gravitational potential and thus produces stronger shock-heating and greater mass loss (e.g.~\citealt{Bauswein+13}).

\begin{table}
\caption{Sources of $r$-Process Ejecta in Binary Neutron Star Mergers \label{table:ejecta}}

\begin{tabular}{ccccccc}

Ejecta Type & $M_{\rm ej} (M_{\odot})$ & $v_{\rm ej}(c)$ & Color & $M_{\rm ej}$ decreases with & References \\

\hline

Tidal Tails & $\sim 10^{-4}-10^{-2}$ & $0.15-0.35$ & Red (NIR) & $q = M_{2}/M_{1}$ & e.g., 1,2 &  \\
Polar Shocked & $\sim 10^{-4}-10^{-2}$ & $0.15-0.35$ & Blue (visual) & $M_{\rm rem}/M_{\rm max}, R_{\rm ns}$ & e.g., 3$-$5 \\
Disk Outflows & $10^{-4}-0.07$ & $0.03-0.1$ & Blue+Red & $M_{\rm rem}/M_{\rm max}$ & e.g., 6$-$8 \\
\hline \\
\end{tabular}
\\
(1) \citealt{Rosswog+99}; (2) \citealt{Hotokezaka+13}; (3) \citealt{Bauswein+13}; (4) \citealt{Sekiguchi+16}; (5) \citealt{Radice+16}; (6) \citealt{Fernandez&Metzger13}; (7) \citealt{Perego+14}; (8) \citealt{Just+15}; (9) \citealt{Siegel&Metzger17}
\end{table}

Debris from the merger which is not immediately unbound can possess enough angular momentum to circularize into an accretion disk around the central remnant, providing an agent to power an ultra-relativistic GRB jet (e.g.~\citealt{Narayan+92}; $\S\ref{sec:GRB}$).  Slower expanding outflows from this remnant disk, which occur on timescales of up to seconds post merger, provide another important source of $r$-process ejecta (e.g.~\citealt*{Metzger+08}; \citealt{Dessart+09}).  The quantity of mass in the disk outflows $M_{\rm ej}^{\rm disk}$ scales approximately with the original mass of the torus $M_{\rm t}$, with $M_{\rm ej}^{\rm disk} \approx 0.2-0.4M_{\rm t}$ \citep{Fernandez&Metzger13,Perego+14,Just+15,Fernandez+15,Siegel&Metzger17}.  Because the mass of the torus increases with both the mass ratio of the binary and the lifetime of the hypermassive NS (e.g.~\citealt{Hotokezaka+13}), $M_{\rm ej}^{\rm disk}$ is also an decreasing function of $q$ and $M_{\rm rem}/M_{\rm max}$, i.e. an asymmetric merger or long-lived NS remnant produces a greater quantity of disk ejecta.  For a massive torus $\gtrsim 0.1-0.2M_{\odot}$ the disk ejecta mass $M_{\rm ej}^{\rm disk} \sim 0.05-0.1M_{\odot}$ can greatly exceed that of the dynamical ejecta.  The electron fraction distribution of the disk outflows, though generally broad $Y_{e} \sim 0.1-0.4$ \citep{Just+15}, depends on the lifetime of the central neutron star remnant due to the de-neutronizing impact of its strong electron neutrino luminosity.  The average $Y_e$ of the disk outflow grows with the time the hyper- or supra-massive NS survives before collapsing to a BH  (\citealt{Metzger&Fernandez14,Perego+14}).  Disk outflow simulations find that the unbound matter achieves asymptotic speeds $v_{\rm ej} \approx 0.03-0.1$ c which are typically $2-3$ times lower than the velocity of the dynamical ejecta.

Figure \ref{fig:ejecta} and Table \ref{table:ejecta} summarizes the quantity of lanthanide-poor (``blue") and lanthanide-rich (``red") ejecta from neutron star mergers from both dynamical and disk wind channels, and their dependence on properties of the binary (remnant mass $M_{\rm rem}$, mass ratio) and neutron star (radius, maximum $M_{\rm max}$).  The disk wind ejecta exhibits a complex behavior with increasing remnant lifetime (decreasing $M_{\rm rem}/M_{\rm max}$), as outlined schematically by the region between the black solid arrows.

\subsection{Kilonova Emission Models}

\label{sec:KN}

Table \ref{table:KN} summarizes the historical progression of kilonova models and their predictions for the luminosity, timescale, and color of the thermal emission.  Although models capable of explaining the detailed color evolution of the emission following~\LIGO~reached their present mature form just in the last couple of  years, many of the basic predictions were in place earlier.

\citet{Li&Paczynski98} first proposed that the radioactive ejecta of a NS merger could power a supernova-like thermal transient.  Due to the small quantity of ejecta mass and its high expansion speed $\sim 0.1$ c, they predicted that the ejecta would become diffusive to photon radiation and the emission would peak on a timescale of $\lesssim 1$ day, much shorter than the rise time of a supernova.  However, \citet{Li&Paczynski98} did not possess a physical model for radioactive heating rate $\dot{q}$ (e.g.~based on a nuclear reaction network; the term ``$r$-process" does not appear in their paper), which they instead parameterized as $\dot{q} \propto t^{-1}$ with the normalization left as a free parameter.  Since the peak luminosity is proportional to the heating rate at the time of peak light, their fiducial model reached extremely high values $\sim 10^{44}$ erg s$^{-1}$, close to the brightest supernovae ever discovered, with a spectral peak in the ultra-violet.  Such luminous transients following NS merger were disfavored based on observations ruling out their presence following short duration GRBs, after they began to be localized to sufficient accuracy for optical follow-up by the {\it Swift} satellite starting in 2005 (e.g.~\citealt{Fox+05,Hjorth+05,Berger+05,Bloom+06}).

\citet{Metzger+10} first calculated the late-time radioactive heating from decaying $r$-process nuclei (predicting $\dot{q} \propto t^{-1.3}$ on timescales of hours to days; Fig.~\ref{fig:Lbol}), which they incorporated self-consistently into the light curve calculation.  They also used a more physical model for the opacity, assuming it was provided by the line (bound-bound) opacity of iron versus the (highly sub-dominant) electron scattering opacity assumed in earlier work.  They predicted peak luminosities of $\sim 3\times 10^{41}$ erg s$^{-1}$ for $10^{-2}M_{\odot}$ of ejecta expanding at $v \sim 0.1$ c and a spectral peak at visual wavelengths.  As this was roughly 1000 times more luminous than classical novae (which peak typically close to the Eddington luminosity of $\sim 10^{38}$ erg s$^{-1}$), they dubbed these events ``kilonovae."  \citet{Metzger&Berger12} highlighted that the isotropic nature of the kilonova emission, as compared to the tightly collimated and relativistically beamed GRB/afterglow emission, would make them the most promising counterpart for a typical binary NS merger at 200 Mpc, the range of Advanced LIGO/Virgo at design sensitivity.  \citet{Kasliwal&Nissanke14} emphasized that during the few few observing runs with Advanced LIGO, mergers could occur much closer than 200 Mpc and thus kilonovae could be detected even with 1 m class telescopes (as turned out to be the case for \LIGO). 

\citet{Kasen+13} and \citet{Barnes&Kasen13} and subsequently, \citet{Tanaka&Hotokezaka13}, performed the first kilonova calculations including line opacity data based on atomic data expected for ejecta containing heavy $r$-process elements.  They showed that if the ejecta contains lanthanide or actinide nuclei with partially-filled $f$-shell valence electron shells, as occurs if the $r$-process passes the second abundance peak at atomic mass number $A \approx 130$, then the resulting photon opacity at UV and optical wavelengths is $\gtrsim 10-100$ times greater than if the ejecta were composed of iron-like nuclei with partial d-shell valence electrons.  This high optical opacity delays the evolution timescale of the light curve from $\sim 1$ day to $\sim 1$ week and pushes the spectral peak from visual frequencies predicted by \citet{Metzger+10} and \citet{Roberts+11} to near-infrared wavelengths \citep{Barnes&Kasen13,Tanaka&Hotokezaka13,Grossman+14,Wollaeger+17}.  
%Excess NIR emission observed on timescales of $\sim 1$ week following a short GRB \citep{Tanvir+13,Berger+13}, consistent with the expectations of high-opacity kilonova emission. 

\begin{table}
\caption{Historical Development of Kilonova Models \label{table:KN}}

\begin{tabular}{ccccccc}
\hline
Model & Opacity Source & $L_{\rm peak}$ (ergs s$^{-1}$) & $t_{\rm peak}$ & SED Peak & Ref. \\

\hline
\hline

parameterized heating & e-scattering & $10^{43}-10^{44}$ & $\sim$ 1 day & UV & 1 \\
%$^{56}$Ni heating  & e-scat & $10^{41}-10^{42}$ & $\sim 1$ day & visual & 2 \\ 
$r$-process heating & iron & $10^{41}-10^{42}$ & $\sim 1$ day & visual & 2, 3 \\
La Opacities & heavy $r$ & $10^{40}-10^{41}$ & $\sim 1$ week & NIR & 4, 5 \\
``Blue" + ``Red" & light + heavy $r$ & $10^{40}-10^{42}$ & $1$ day $\rightarrow$ 1 week & visual$\rightarrow$NIR & e.g., 6, 7, 8 \\
\hline 
\hline \\
\end{tabular}
\\ (1)  \citealt{Li&Paczynski98}; (2)  \citealt{Metzger+10}; (3) \citealt{Roberts+11}; (4) \citealt{Barnes&Kasen13}; (5) \citealt{Tanaka&Hotokezaka13}; (6) \citealt{Metzger&Fernandez14}; (7) \citealt{Martin+15}; (8) \citealt{Tanaka17}
\end{table}

Although lanthanide opacities move the kilonova emission to the infrared, {\bf not all portions of the NS merger ejecta necessarily produce such heavy nuclei}.  In particular, ejecta with $Y_{e} \gtrsim 0.25$ lacks sufficient neutrons for neutron-capture reactions to push the nuclear flow past the second $r$-process peak at $A \approx 130$ \citep{Metzger&Fernandez14}.  In such a case, the lanthanides are not produced, and the ejecta would produce a ``blue" and fast-evolving kilonova similar to original expectations \citep{Metzger+10} because the opacity of light $r$-process nuclei is only moderately higher than that of iron \citep{Tanaka17,Kasen+17}. 

At least a small quantity of ejecta with $Y_{e} \lesssim 0.2$ will be present in any merger from the tidal tail ejecta or disk winds, making ``red" kilonova emission a ubiquitous feature.  However, outflows from the accretion disk are more isotropic and thus should expand also into the lanthanide-poor polar regions \citep{Metzger&Fernandez14,Perego+14,Martin+15}, powering a separate component of ``blue" emission similar to original kilonova models \citep{Metzger+10}.   In original hybrid ``blue" + ``red" scenarios, the quantity of red versus blue kilonova emission originates from the disk outflows and is diagnostic of the lifetime of the central merger remnant (Fig.~\ref{fig:ejecta}; \citealt{Metzger&Fernandez14,Martin+15}).  However, \citet{Wanajo+14} showed that when neutrino transport effects are included, the shock-heated polar dynamical ejecta may also be lanthanide-free with $Y_{e} \gtrsim 0.25$ (cf.~\citealt{Goriely+15,Sekiguchi+16}), in which case it could also contribute to$-$or even dominate$-$the early-time blue kilonova emission.  As depends on the relative velocity of the red and blue ejecta, the blue emission is visible only for viewing angles along which it is not blocked by the higher opacity red matter (e.g.~\citealt{Kasen+15}).

\subsection{Interpreting the Kilonova Which Accompanied \LIGO}

\label{sec:KNinterp}

The thermal spectrum of the optical counterpart of \LIGO~(e.g.~\citealt{Nicholl+17,Chornock+17,Levan&Tanvir17}) strongly supported the kilonova model, as compared to the power-law spectrum expected for non-thermal GRB afterglow emission.  The shape of the bolometric light curve following peak is broadly consistent with the $\propto t^{-1.3}$ radioactive heating rate from freshly synthesized $r$-process nuclei (Fig.~\ref{fig:Lbol}; \citealt{Metzger+10}).  Over the first few days the transient colors were blue and rapidly-evolving with a spectral peak at visual wavelengths (e.g.~\citealt{Soares-Santos+17,Smartt+17,Troja+17,Nicholl+17,Evans+17,Shappee+17,Pian+17,McCully+17,Cowperthwaite+17}).  At later times, the colors became substantially redder and more slowly-evolving on timescales of several days to a week, with a spectral peak around 1.5$\mu$m (e.g.~\citealt{Chornock+17,Tanvir&Levan17,Pian+17,Shappee+17,Kasliwal+17}).  The lack of well-defined spectral features is suggestive of line blending due to the photosphere expanding at speeds up to several tenths of the speed of light \citep{Nicholl+17}, though broad undulations in the NIR spectra predicted from lathanide absorption \citep{Kasen+13} were possibly observed in \LIGO~(e.g.~\citealt{Chornock+17,Troja+17}).  As pointed out by several works (e.g.~\citealt{Kasen+17,Cowperthwaite+17,Chornock+17,Kasliwal+17,McCully+17,Smartt+17,Troja+17,Pian+17,Arcavi+17b}), these properties are consistent with the two-component blue+red kilonova picture discussed above (e.g.~\citealt{Metzger&Fernandez14,Tanaka+17}).      

{\bf What part of the merger or its aftermath created the ejecta we observe?}  Both the dynamical merger and the subsequent accretion disk wind can contribute to the ejecta ($\S\ref{sec:rprocess}$, Table \ref{table:ejecta}), making it important to carefully assess the origin of the dominant contribution to the blue and red ejecta components (see \citealt{Kasen+17}, \citealt{Cowperthwaite+17} for a complementary discussion).  The quantity of blue (lanthanide-free) ejecta from \LIGO~was estimated to be $\approx 1-2\times 10^{-2}M_{\odot}$ with a mean velocity of $v_{\rm ej} \approx 0.2$ c \citep{Cowperthwaite+17,Nicholl+17}, based on fitting the observed light curves to kilonova models \citep{Metzger17} and the spectra to more detailed radiative transfer calculations \citep{Kasen+17}.  Comparing these measurements to the results of numerical simulations ($\S\ref{sec:rprocess}$), the high velocity tentatively supports an origin associated with the shock-heated dynamical ejecta \citep{Wanajo+14,Goriely+15,Sekiguchi+16} rather than a disk wind.  In such a case, the large {\it quantity} of the dynamical ejecta would suggest the radii of the merging NSs were relatively small $\lesssim 11$ km \citep{Nicholl+17}.  If confirmed by additional modeling and numerical simulation work, this result would have key implications for the equation of state of the NS (e.g.~\citealt{Ozel&Freire16}).    

The total mass of the red (lanthanide-bearing) ejecta was estimated to be $\approx 4\times 10^{-2}M_{\odot}$ with a somewhat lower expansion velocity $v \approx 0.1$ c than the blue ejecta (e.g.~\citealt{Cowperthwaite+17,Chornock+17,Nicholl+17}).  Such a large quantity of ejecta, if originating from the tidal tidal tail, would require an extremely asymmetric merger (e.g.~\citealt{Hotokezaka+13}); however, this would not explain the low ejecta velocity, which based on numerical simulations is expected to be $\approx 0.2-0.3$ c.  Accretion disk winds provide a more natural explanation, as several $10^{-2}M_{\odot}$ expanding at $v \approx 0.1$ c matches theoretical expectations for the outflow from a massive torus $\gtrsim 0.1M_{\odot}$ (e.g.~\citealt*{Metzger+08}; \citealt{Fernandez&Metzger13,Just+15,Wu+16,Siegel&Metzger17}).  A relatively spherical accretion disk outflow could be consistent with the non-detection of linear polarization from the late red kilonova emission \citep{Covino+17}.

Such a large torus is not expected if the merger event resulted in the prompt collapse to a BH, but instead suggests that at least a temporarily-stable hypermassive NS formed in \LIGO~(e.g.~\citealt{Shibata&Taniguchi06}).  On the other hand, the fact that the disk outflows produced primarily ejecta with $Y_{e} \lesssim 0.25$ (as needed to power red kilonova emission) would, based on the results of numerical simulations of the disk evolution \citep{Lippuner+17}, implicate a relatively short hypermassive NS lifetime, $\lesssim 100$ ms.  This is consistent with the moderate inferred kinetic energy of the kilonova ejecta $\approx 10^{51}$ erg, which does not require substantial additional energy input from the rotational energy of a temporarily-stable supramassive NS remnant \citep{Metzger&Piro14,Margalit&Metzger17}.  

The kilonova emission from \LIGO~probes the merger ejecta structure for one particular viewing angle and set of initial binary parameters.  Future NS mergers observed from different viewing angles, or with a different total binary mass or binary asymmetry, could produce a quantitatively or qualitatively different signal.  Some of these possibilities are described in the final take-aways section ($\S\ref{sec:takeaways}$).

%These dual constraints$-$to avoid prompt collapse, but also not form a long-lived stable remnant$-$can be combined to constrain the maximum mass of the neutron star \citep{Margalit&Metzger17,LIGO+17PARAMS}.  For \LIGO \citep{Margalit&Metzger17} that the latter constraint is tighter, giving $M_{\rm max} \lesssim 2.25M_{\odot}$ at 90\% confidence. 

{\bf How uncertain are the ejecta masses?}  Uncertainties enter estimates of the kilonova ejecta mass from at least three sources: geometry, the radioactive heating rate, and the thermalization efficiency of the decay products.  While the thermalization efficiency of the early-time blue kilonova is relatively robust \citep{Metzger+10}, the total radioactive heating rate of $Y_{e} \gtrsim 0.25$ matter is uncertain at the factor of a few level \citep{Lippuner&Roberts15}.  By contrast, while for the red kilonova emission the total radioactive heating rate of $Y_{e} \lesssim 0.25$ matter is robust to within a factor $\lesssim 2$ \citep{Metzger+10,Korobkin+12}, the thermalization efficiency is less certain because it depends on the distribution of heating between $\beta-$decays, $\alpha$-particle decay and fission \citep{Barnes+16,Rosswog+17}, which depend on the unknown masses of nuclei well off the valley of nuclear stability \citep{Mendoza-Temis+15,Hotokezaka+15}.  Geometric effects also typically enter at the factor $\approx 2$ level \citep{Roberts+11,Grossman+14,Kasen+17}.  A reasonable guess is that the blue kilonova ejecta mass is accurate to a factor of $\approx 2-3$, but the red KN ejecta mass could be uncertain to a factor of $\approx 3-10$.  Even given these uncertainties, and those on the overall rate of binary NS mergers, the discovery of \LIGO~makes it likely that binary NS mergers are the dominant site of $r$-process nuclei in the universe \citep{Lattimer&Schramm74,Symbalisty&Schramm82,Eichler+89,Freiburghaus+99}.

%\begin{figure}[!t]
%\includegraphics[width=0.5\textwidth]{MF14.png}
%\hspace{0.0cm}
%\caption{\footnotesize Two-component (``blue" + ``red") kilonova model from \citet{Metzger&Fernandez14}.  Near-infrared emission from lanthanide-rich components of the tidal ejecta peaking on a timescale of $\sim 1$ week \citep{Barnes&Kasen13,Tanaka&Hotokezaka13} should be visible from any angle, while visual emission from lanthanide-free ejecta \citep{Metzger+10} may be visible along the polar direction.  Polar ejecta may originate from the accretion disk outflows \citep{Metzger&Fernandez14}, especially in the case of a long-lived HMNS \citep{Perego+14,Martin+15,Lippuner+17}.  Contemporary numerical simulations of binary NS merger have shown that high $Y_{e}$ (lanthanide-free) matter may also be ejected dynamically in the polar regions \citep{Wanajo+14,Goriely+15,Sekiguchi+16,Radice+16}.     }
%\label{fig:Lbol}
%\end{figure}

\section{A Gamma-Ray Burst and an Afterglow}
\label{sec:GRB}

\LIGO~was accompanied by a short burst of gamma-rays (\citealt{Goldstein+17,Savchenko+17,LIGO+17FERMI}, dubbed GRB170817A), which was similar in duration to$-$but orders of magnitude less energetic than$-$standard cosmological short GRBs (see also \citealt{Fong+17}).  The onset of GRB170817A was delayed by $\approx$1.7 seconds relative to the end of the merger, as inferred from the GW signal.  This near temporal coincidence enabled constraints to be placed on fundamental physics, such as the difference between the speed of EM and gravitational waves \citep{LIGO+17FERMI}.  Given the inference described above from the red kilonova emission that a massive accretion disk formed and that BH formation was relatively prompt in \LIGO, such a torus-BH system provides a natural engine for powering a relativistic GRB jet (e.g.~\citealt{Narayan+92,Aloy+05}).  

GRB170817A was composed of two separate emission components, (1) an initial hard spike lasting $\lesssim 0.5$ seconds with a non-thermal spectrum broadly similar to normal short GRBs, followed by (2) a softer emission component lasting a few seconds with a spectrum consistent with being thermal (\citealt{Goldstein+17,Savchenko+17}).  As discussed by \citet{LIGO+17FERMI}, the first emission component could be the off-axis signature of a much more powerful short GRB jet, which either has its luminosity de-boosted by relativistic beaming or, alternatively, is ``structured" in polar angle or time (e.g.~\citealt{Lamb&Kobayashi16,Kathirgamaraju+17}).  A thermal component could originate from the hot cocoon \citep{Lazzati+17,Nakar&Piran17} or shock break-out \citep{Nakar&Sari12} created as the ultra-relativistic GRB drills through the polar merger ejecta cloud (e.g.~\citealt{Duffell+15}).  

Temporal evolution of the jet's structure is also natural.  The post-merger accretion disk evolves on the viscous time of several seconds, over which time its mass is substantially depleted by accretion and outflows (e.g.~\citealt{Fernandez&Metzger13}; $\S\ref{sec:rprocess}$).  If disk winds and the dynamical ejecta are the medium responsible for collimating the GRB jet then$-$as the density of the surrounding ejecta cloud  and the jet power weaken in time$-$the jet opening angle may also widen on a timescale of a few seconds, similar to the observed delay of GRB170817A.\footnote{Such evolution in the jet properties would not necessarily be expected in long duration GRBs because the outer mantle of the progenitor star responsible for jet collimation in this case does not evolve appreciably on the timescale of the burst.}
  
\begin{SCfigure}
\centering
\includegraphics[width=0.6\textwidth]
{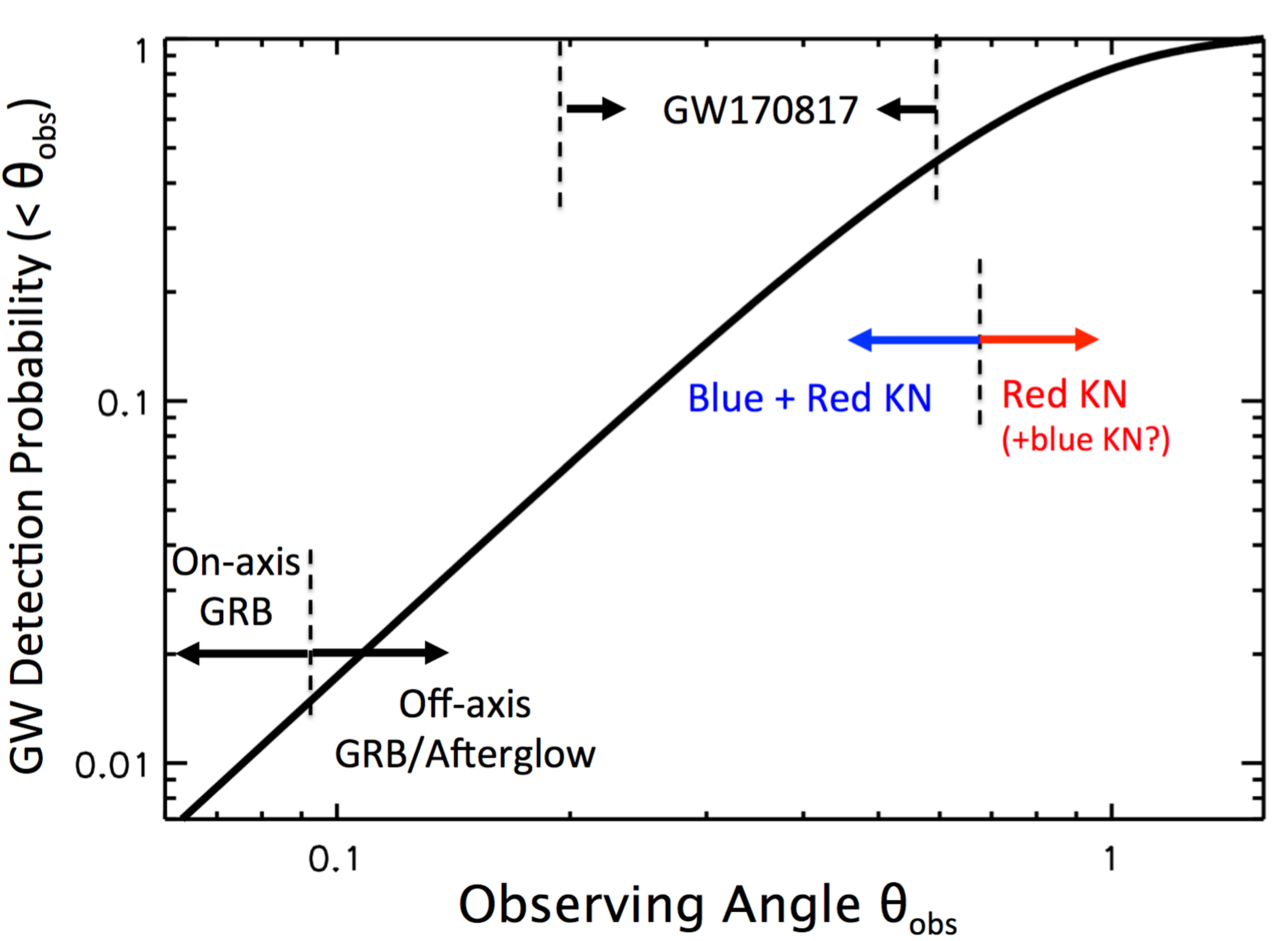}
% un-comment the following line to include your fig1b.ps postscript file:
\caption{\footnotesize Probability of detecting GWs from a binary NS merger at an observing angle (as measured from the binary orbital axis) less than a value $\theta_{\rm obs}$ from \citet{Schutz11}.  Shown for comparison are (1) the $1\sigma$ range of inclination angles inferred for \LIGO~(when the degeneracy with distance is broken by the host galaxy distance; \citep{LIGO+17H0}); (2) the approximate angle separating viewers within the lanthanide-free polar funnel of the dynamical ejecta \citep{Sekiguchi+16}, which observe both a blue and red kilonova signature, from those more equatorial viewers who might observe only the red kilonova component (if the blue kilonova is obscured); (3) the approximate angle $\theta_{\rm obs} \lesssim \theta_{\rm j} \approx 0.05-0.2$ separating mergers which produce on-axis cosmological short GRBs with prompt X-ray afterglows, from mergers viewed outside the jet axis who instead observe a weaker GRB and delayed orphan afterglow emission, as may have been observed in \LIGO.  Adapted from a similar figure in \citet{Margutti+17}.}
\label{fig:angle}
\vspace{-0.2cm}
\end{SCfigure}

The possibility that \LIGO~was viewed off-axis angle relative to the core of the GRB jet is consistent with the relatively large binary inclination angle relative to our line of site, $\theta_{\rm obs} \approx 0.2-0.6$ (Table \ref{table:BNS}).  The local rate of short GRBs viewed on-axis is $f_{\rm on}\mathcal{R}_{\rm SGRB} \approx 2-6$ Gpc$^{-3}$ yr$^{-1}$ \citep{Wanderman&Piran15} is much less than the total GW-inferred rate of binary NS mergers
$\mathcal{R}_{\rm BNS} \approx 1540^{3200}_{-1220}$Gpc$^{-3}$yr$^{-1}$ \citep{LIGO+17DISCOVERY}.  Here $f_{\rm on} \approx 0.3-1$ is the fraction of detected short GRBs which are observed on-axis; at a minimum this equals the $\approx 30\%$ of short GRBs with prompt and luminous X-ray afterglows detected by {\it Swift} that also enable the burst to be well-localized and the host galaxy to be identified to cosmological distances $z \sim 0.1-1$ (e.g.~\citealt{Fong+15}).  Assuming that every binary NS merger produces a short GRB, and that all short GRBs are binary NS mergers, the implied beaming fraction $f_{\rm b} \approx f_{\rm on}\mathcal{R}_{\rm SGRB}/\mathcal{R}_{\rm BNS} \sim 1\times 10^{-4}-2\times 10^{-2}$ and a corresponding half-opening angle for the core of the GRB jet of $\theta_{\rm j} = (2f_{\rm b})^{1/2} \approx 0.02-0.2$.  Thus, we infer that \LIGO~was indeed most likely viewed outside the core of the jet, such that $\theta_{\rm obs}/\theta_{\rm j} \approx 1-30$ (see Fig.~\ref{fig:angle}).  

Additional evidence suggesting the presence of a more powerful off-axis jet is the discovery of non-thermal X-ray \citep{Margutti+17,Troja+17} and radio emission \citep{Hallinan+17,Murphy+17,Mooley+17,Alexander+17} following the merger with a delayed rise of several weeks.  Such emission is naturally expected from an off-axis ``orphan" GRB afterglow (e.g.~\citealt{Granot+02}).  At earlier times, the afterglow emission was relativistically beamed away from our line of site; however, as the GRB ejecta sweeps up gas from the ISM of the surrounding galaxy of density $n$, it begins to decelerate such that its Lorentz factor approaches a self-similar evolution with radius $r$ \citep{Blandford&McKee76},
\be
\Gamma =  \left(\frac{17}{8\pi}\frac{E_{\rm j}}{n m_{\rm p}r^{3}\theta_{\rm j}^{2}c^{2}}\right)^{1/2},
\ee
where $E_{\rm j}$ is the total beaming-corrected energy of the jet.  The initially de-beamed emission enters the our causal cone, and the off-axis afterglow peaks, once $\Gamma \approx 1/\theta_{\rm obs}$.  Using the relationship between emission radius and observer time $t = r/(2\Gamma^{2}c)$, this occurs on a timescale
\be
t_{\rm pk} \approx \left(\frac{17 E_{\rm j}}{64\pi n m_p c^{5}}\frac{\theta_{\rm obs}^{8}}{\theta_{\rm j}^{2}}\right)^{1/3} \approx 18\,\,{\rm days}\,\left(\frac{n}{0.01\,{\rm cm^{-3}}}\right)^{1/3}\left(\frac{E_{\rm j}}{10^{50}\,{\rm erg}}\right)^{1/3}\left(\frac{\theta_{\rm obs}}{5\theta_{\rm j}}\right)^{2/3}\left(\frac{\theta_{\rm obs}}{0.3}\right)^{2},
\ee
where we have used the relationship between radius and the photon arrival time $t = r/(2\Gamma^{2}c)$.  Matching the estimated peak timescale of the X-ray and radio emission of $\approx 15-30$ days (e.g.~\citealt{Margutti+17,Troja+17,Hallinan+17,Alexander+17}) suggests jet energies $E_{\rm j} \sim 10^{49}-10^{50}$ erg and ISM densities $n \sim 10^{-4}-10^{-2}$ cm$^{-1}$ (e.g.~\citealt{Margutti+17,Alexander+17,Fong+17}), broadly consistent with those inferred for on-axis short GRB \citep{Nakar07,Berger14} and much less than those of star-forming environments which characterize long-duration GRBs.

Most of the $\approx 75\%$ of short GRBs discovered by {\it Swift} which are accompanied by luminous X-ray afterglows are presumably those events viewed on-axis, within or nearly within the opening angle of the jet, $\theta_{\rm obs} \lesssim \theta_{\rm j}$.  However, it is interesting to ask what fraction of the remaining 25\% could be off-axis dim nearby bursts similar to that observed from \LIGO~and located at much closer distances $\approx 40-100$ Mpc (see also \citealt{LIGO+17FERMI}).  The fraction of all mergers that would be viewed at the inferred inclination angle of~\LIGO~is $\approx (\theta_{\rm obs}/\theta_{\rm j})^{2} \sim 10-100$ times larger than the on-axis events.  However, the isotropic fluence of the GRB associated with \LIGO~of $E_{\rm iso} \sim 4\times 10^{46}$ erg \citep{Goldstein+17} was $\sim 10^{2.5}$ times lower than the dimmest values for the cosmological short GRB population, in which case the detection volume is smaller by a factor of $\gtrsim 10^{4}$.  Though crude, this estimate suggests that less than a few percent of the total short GRB population could arise from NS mergers as close as \LIGO.  A relatively nearby population (analogs of \LIGO) appears to be broadly consistent with the inference by \citet{Tanvir+05} that $\approx 10-25\%$ of short GRBs originate from $\lesssim 70$ Mpc, based on a correlation between the sky position of BATSE short GRBs with local structure defined by catalogs of local galaxies.

\section{Lessons Learned and Open Questions}
\label{sec:takeaways}

Taking at face value the unified scenario for the multi-wavelength counterparts to \LIGO~summarized by Fig.~\ref{fig:schematic}, I now recount what I believe are the major take-away lessons from the first binary NS merger~\LIGO~with EM follow-up.  I address the question of how ``typical" we should expect the EM signal from \LIGO~to be, and, conversely, what counterpart diversity is expected as we move ahead to the era in which LIGO/Virgo approach design sensitivity and the NS mergers may be detected as frequently as once per week.

\begin{itemize}

\item{{\bf A triumph for theory.}  While the detection of gamma-ray emission from an off-axis jet was  surprising to many (however, see \citealt{Lazzati+17}), perhaps the biggest take-away from \LIGO~is that theorists predicted the observed EM signals more or less accurately; the merger was surprisingly well-behaved.  The discovery of both blue \citep{Metzger+10} and red \citep{Barnes&Kasen13,Tanaka&Hotokezaka13} kilonova emission was observed with the timescale, luminosity (indeed, $\sim 1000$ times brigher than a classical nova!), and colors predicted by theory (Fig.~\ref{fig:Lbol}) for an ejecta mass and velocity consistent with those predicted by numerical simulations of the merger (e.g.~\citealt{Rosswog+99,Oechslin&Janka06,Sekiguchi+16}) and the post-merger accretion flow (e.g.~\citealt{Fernandez&Metzger13}).   Likewise, the prediction for an off-axis orphan afterglow, though not without degeneracies or free parameters, are broadly consistent with the observed non-thermal X-ray and radio emission for an observer situated at an angle relative to the binary axis \citep{Granot+02,VanEerten+10}.  This provides the most direct evidence yet that binary NS mergers are the source of most or all classical short GRBs observed at cosmological distances (e.g.~\citealt*{Eichler+89}).}

\item{{\bf An abundance of riches.}  Given the significant quantity of $r$-process nuclei produced in \LIGO, along with the relatively high implied merger rate, this strongly supports binary NS mergers as the dominant source of heavy $r$-process nuclei in the galaxy (e.g.~\citealt{Kasen+17,Cowperthwaite+17,Chornock+17,LIGO+17kilonova}), confirming long-standing theoretical ideas \citep{Lattimer&Schramm74,Symbalisty&Schramm82,Eichler+89,Freiburghaus+99}.  The red kilonova emitting ejecta component dominates the total ejecta mass and thus likely also dominates the yield of both light and heavy $r$-process nuclei (Table \ref{table:BNS}).  Assuming an $r$-process abundance pattern matching the solar one, one infers that over a hundred Earth masses in gold was created within a few seconds following \LIGO.  }

%However, light $r$-process nuclei are actually more common than heavy $r$-process nuclei in our solar system by ratio of light:heavy of 5:1, inconsistent with the opposite 3:1 heavy:light ratio inferred for \LIGO~(Table \ref{table:BNS}).  Though more detailed modeling of the observed kilonova emission may ultimately relieve this tension, this appears to leave open the possibility that additional astrophysical channels contribute to the abundances of the light $r$-nuclei, such as rare supernovae  (\citealt{Thompson04,Metzger+07,Winteler+12,Mosta+14}).  Alternatively, \LIGO~might not be representative of the entire merger population, e.g.~if the light $r$-process yield is higher in mergers for which BH formation is not as prompt (e.g.~\citealt{Lippuner+17}).}  

\item{{\bf Similar event, but different viewing angle?}  \LIGO's relatively face-on orientation of $\theta_{\rm obs} \approx 0.2-0.6$ will be shared by only $\approx 10-50\%$ of GW-discovered mergers (Fig.~\ref{fig:angle}).  For larger inclination angles, theory suggests the GRB and afterglow emission are unlikely to be sufficiently luminous to be detected, though observationally this is not yet well constrained (\citealt{LIGO+17FERMI}).  The kilonova is predicted to be relatively isotropic \citep{Roberts+11} and thus in principle should be visible with a similar luminosity (to within a factor of $\approx 2$) for viewers observing the event closer to the binary plane.  However, if the speed of the high-opacity lanthanide-rich equatorial tidal matter exceeds the speed of the blue polar ejecta, then the blue emission could be blocked or at least partially suppressed for the roughly half of the mergers viewed at $\theta_{\rm obs} \gtrsim 0.6$ \citep*{Kasen+15}.  Although the average velocity of the red ejecta from \LIGO~was inferred to be less than for the blue ejecta, even a small quantity $\lesssim 10^{-3}M_{\odot}$ of the lanthanide-rich matter moving at high velocities (which is challenging to rule out observationally) would be enough to at least partially attenuate the blue emission for less polar viewing angles.  Much will be learned about the geometry of the kilonova ejecta from the luminosity of the early blue emission for the next NS merger observed at a higher inclination angle, closer to within the binary plane.   }

\item{{\bf Similar event, but greater distance?}  Luck may have played some role in the first NS merger discovery occurring at only 40 Mpc.  For the same merger viewed at a distance more typical of those expected during LIGO/Virgo's O3 science run next fall ($\gtrsim 100$ Mpc), the gamma-ray, X-ray and radio luminosities observed in \LIGO~are probably too dim to be detected with current facilities.  By contrast, the early blue kilonova observed on timescales of $\lesssim 1$ day would still reach a visual magnitude of $R = 19.5$ at 100 Mpc or $R = 21$ at 200 Mpc, within the reach of moderate-sized wide-field follow-up telescopes, such as the Zwicky Transient Facility \citep{Bellm14} and the BlackGEM array \citep{Bloemen+16}.  Even kilonovae for which the early blue emissoin is blocked or suppressed would be detectable to 200 Mpc distances by more sensitive telescopes such as DECam (\citealt{Cowperthwaite+17b}b).  With a magnitude depth of $R = 25-26$, the Large Synoptic Survey Telesope (LSST) could detect a similar blue kilonova to distances $\gtrsim 1$ Gpc!  Kilonovae still represent the counterpart most likely to accompany the majority of mergers \citep{Metzger&Berger12}.}

\item{{\bf Similar event, but greater binary mass?}  The large quantity of ejecta from \LIGO~suggests that a temporarily stable hypermassive NS remnant was created during the merger.  The total binary mass $\approx 2.3-2.78M_{\odot}$ (Table \ref{table:BNS}, assuming low NS spin) is broadly consistent with that expected by drawing two stars from the the Galactic NS population (well-described by a Gaussian of mean $\mu = 1.32M_{\odot}$ and standard deviation $\sigma = 0.11M_{\odot}$; \citealt{Kiziltan+13}).  For more massive binaries (with precisely how massive$-$and thus how rare$-$depending sensitively on the maximum NS mass), a prompt collapse would occur instead of the formation of a hypermassive NS.  In such a case the blue kilonova would be strongly suppressed, especially the polar dynamical component.  The red kilonova also might now be dominated by the tidal tail ejecta instead of the disk outflows and thus would also be somewhat dimmer.  All else being equal (e.g. similar observing inclination), an inverse relationship is expected between the kilonova luminosity and the total binary mass.}

\item{{\bf Similar event, but lower binary mass?}  For less massive binaries, a long-lived supramassive or stable NS remnant could form instead of a short-lived hypermassive NS (e.g.~\citealt{Ozel+10,Piro+17}).  In such cases, a fraction of the substantial rotational energy of the remnant (communicated, e.g., by magnetic spin-down) is likely to be transferred to the kilonova ejecta \citep{Bucciantini+12}, accelerating it to higher speeds $v \sim c$ than inferred for \LIGO.  This additional source of acceleration and rotational energy input may power a signal with an optical and X-ray luminosity much higher than what is possible from $r$-process heating alone \citep{Yu+13,Metzger&Piro14,Siegel&Ciolfi16} and a faster evolution timescale.  An ultra-relativistic outflow from a long-lived remnant may also power the mysterious variable X-ray/gamma-ray emission observed for hundreds or thousands of seconds following some short GRBs \citep{Norris&Bonnell06,Rowlinson+13}.  No evidence for such ``extended emission" was observed at hard X-ray/gamma-ray energies following \LIGO~\citep{LIGO+17FERMI}.  Still, it would not be surprisingly to see qualitatively different EM behavior from the low mass tail of the NS binary merger population.}

\item{{\bf BH-NS instead of NS-NS?} What if \LIGO~had been the merger between a NS and a stellar mass BH instead of a NS-NS?  If \LIGO's blue kilonova was indeed the result of matter squeezed out of the polar region by the colliding NSs, then a similar component of fast high-$Y_{e}$ ejecta and its concomitant blue kilonova emission will not be present for BH-NS mergers.  If the mass of the BH is sufficiently low, and/or its spin sufficiently high, to disrupt the NS outside of the BH innermost stable circular orbit, then the tidal ejecta mass is typically $\sim 0.1M_{\odot}$ \citep{Foucart12}, much higher than that in NS-NS mergers.  Although this is consistent with the higher red ejecta mass inferred for \LIGO, the average velocity of the tidal ejecta from a NS-BH will be higher, closer to $\approx 0.2-0.3$ c, than inferred for \LIGO.  A disrupted NS will also produce a massive accretion torus as in the NS-NS case, potentially capable of powering a GRB jet.  A moderate quantity of blue kilonova ejecta from the accretion disk outflows are expected in this case (e.g.~\citet{Fernandez+17} finds $\lesssim 8\times 10^{-3}M_{\odot}$ in $Y_e \gtrsim 0.25$ disk wind ejecta; \citealt{Kyutoku+17}).  Again, however, the velocity of the disk outflow is predicted to be lower $\lesssim 0.1$ c than that inferred for \LIGO, and it may be blocked by the tidal ejecta for equatorial viewers.   }  

\item{{\bf Implications for the radii of NSs.}  The inference of a small NS radius $\lesssim 11$ km (e.g.~\citealt{Nicholl+17}) would have critical implications for the equation of state \citep{Ozel&Freire16}.  This conclusion is predicated on the large quantity $\approx 10^{-2}M_{\odot}$ of blue kilonova ejecta inferred from \LIGO~being the result of dynamical ejecta produced by the merger collision, as opposed to outflows from the post-merger accretion disk.  This hypothesis could potentially be checked by searching for similar early blue emission from future NS-NS mergers seen from a similar binary inclination.  The polar dynamical ejecta should be quantitatively similar between NSs of moderately different masses (the radius of a NS is almost independent of its mass for typical masses).  However, the disk outflows should vary more strongly with the total mass of the system and the asymmetry of the binary.  Thus, if the early blue emission of the type observed in \LIGO~is observed with very similar properties for a range of binaries with different masses (and otherwise similar polar viewing angles), this would favor it being dynamical in origin and thus the hypothesis of a small NS radius.  Additional numerical work, with better neutrino transport and resolution at the collision interface between the NSs, are also needed to solidify the use of kilonovae as probes of the NS radius.}

\item{{\bf Implications for the maximum NS mass.}  The lack of evidence for either a prompt collapse (from the large inferred quantity of kilonova ejecta) or the formation of a supramassive NS remnant (from the prompt GRB and moderate kinetic energy of the kilonova ejecta) points to the formation of a hypermassive NS remnant in \LIGO.  Using a suite of representative NS equations of state, the gravitational masses of the NS-NS binary measured by Advanced LIGO/Virgo can be used to place an upper limit on the maximum stable mass of a slowly-rotating NS of $M_{\rm max} \lesssim 2.17M_{\odot}$ at 90\% confidence \citep{Margalit&Metzger17}.  This upper limit on $M_{\rm max}$ is tighter, and arguably less model-dependent, than other observational or theoretical constraints.  This constraint would be strengthened or tightened with the discovery of additional mergers, particularly if the lowest-mass binaries also show no evidence for a supramassive NS remnant.
 }

\item{{\bf Implications for merger cosmology.}  The EM follow-up campaign of \LIGO~demonstrates a proven method to obtain a redshift for a GW event, thus opening the potential of binary NS merger as ``standard sirens" to study the expansion history of the universe \citep{Schutz86,Holz&Hughes05}.  A comparison of the luminosity distance from the GW signal of \LIGO~with the measured redshift of the host galaxy led to a measurement of the Hubble constant of $H_{0} = 69.3^{12.1}_{-6.0}$ km s$^{-1}$ Mpc$^{-1}$ which is completely independent of, and fully consistent with, other estimates of $H_{0}$ \citep{LIGO+17H0}.  The GW detection rate of binary NS mergers will reach 6$-$120 per year once Advanced LIGO/Virgo reach design sensitivity \citep{LIGO+17DISCOVERY}.  A larger sample of GW+redshift events, enabled by kilonova observations, will place increasingly tight constraints on $H_{0}$ \citep{Nissanke+10}.  As knowledge of the angular structure of short GRB jets grows, increasingly realistic priors on the binary inclination may enable the subset of events with detected GRBs of a given luminosity to play an outsized role in the analysis.  

Looking ahead to decade timescales, as additional GW detectors like KAGRA \citep{Aso+13} and LIGO India enter the network and if the Advanced LIGO detectors are upgraded (e.g. to the so-called ``A+" enhanced configuration), then horizon distance approaching $\approx 0.7-2$ Gpc (redshift $z \approx 0.1-0.4$) may be achievable for NS-NS and NS-BH mergers, respectively.  At the latter distances the sample of events with on-axis GRBs also becomes substantial (e.g.~\citealt{LIGO+17FERMI}) and detecting the blue kilonova counterparts, while still possible, would require a telescope with higher sensitivity like LSST.  Once redshifts $z \gtrsim 0.4$ become accessible, GW/EM sirens from NS mergers (particularly the subset of NS-BH mergers with kilonovae and short GRBs) could provide an alternative method to probe additional cosmological parameters such as Dark Energy in a way complimentary to existing methods (e.g.~Ia SNe, Baryon Acoustic Oscillations), but independent of the cosmic distance ladder.  }

\end{itemize}

\section*{Acknowledgements}

This summary is dedicated to Alastair (Al) Cameron (1925-2005).  Cameron discovered the $r$-process in 1957 contemporaneously with B$^{2}$FH and deciphering its astrophysical origin remained a passion throughout his career.  One of Al's last publications, around the time I entered graduate school, hypothesized an $r$-process origin in magnetized jets from compact object accretion disks created in core collapse supernovae \citep{Cameron03}.  Similar magnetized outflows from the post-merger accretion disk (e.g.~\citealt{Siegel&Metzger17}) may have provided the dominant source of the heavy $r$-process inferred from the kilonova emission in~\LIGO~and thus (as far as we can presently discern) the universe as a whole.  I thank Gabriel Martinez-Pinedo for comments on this draft.  I also thank all of my observational collaborators in the DECam GW follow-up team.  BDM was supported in part by NASA grant NNX16AB30G.  

%\bibliographystyle{mn2e}
%\bibliography{refs}

\end{document}